\documentclass[preprint,nofootinbib]{revtex4}%
\usepackage{amssymb}
\usepackage{amsfonts}
\usepackage{amsmath}
\usepackage{amsmath}
\usepackage{graphicx}
\usepackage{tensor}
\usepackage[dvipsnames]{xcolor}  
\providecommand{\U}[1]{\protect\rule{.1in}{.1in}}
\providecommand{\U}[1]{\protect\rule{.1in}{.1in}}
\definecolor{blue}{rgb}{0,0,1}

\definecolor{red}{rgb}{1,0,0}

\usepackage{hyperref}
\hypersetup{
	colorlinks =true,
	linkcolor =NavyBlue,
	citecolor=MidnightBlue,
	filecolor=NavyBlue,
	urlcolor=NavyBlue,
}

\newcommand{\diff}{\mathrm{d}}
\newcommand{\R}{\mathcal{R}}
\newcommand{\lnr}{^{\mathrm{L}}}
\newcommand{\Ls}{L_\star}
\newcommand{\lp}{\ell_{\text P}}
\newcommand{\E}{\mathcal{E}}

\makeatletter
\newcommand{\dal}{\mathop{\mathpalette\dal@\relax}}
\newcommand{\dal@}[2]{%
  \begingroup
  \sbox\z@{$\m@th#1\square$}%
  \dimen0=\fontdimen8
    \ifx#1\displaystyle\textfont\else
    \ifx#1\textstyle\textfont\else
    \ifx#1\scriptstyle\scriptfont\else
    \scriptscriptfont\fi\fi\fi3
  \makebox[\wd\z@]{%
    \hbox to \ht\z@{%
      \vrule width \dimen0
      \kern-\dimen0
      \vbox to \ht\z@{
        \hrule height \dimen0 width \ht\z@
        \vss
        \hrule height 2\dimen0
      }%
      \kern-2.5\dimen0
      \vrule width 2.5\dimen0
    }%
  }%
  \endgroup
}
\makeatother

\begin{document}
\title{Higher-curvature gravity in AdS$_3$, holographic $c$-theorems and black hole microstates}
\author{M. Chernicoff$^1$, G. Giribet$^2$,  
J. Moreno$^3$, J. Oliva$^3$, R. Rojas$^3$, C. Ramirez de Arellano Torres$^3$}
\affiliation{$^1$Departamento de F\'{\i}sica, Facultad de Ciencias, Universidad Nacional Aut\'{o}noma de M\'{e}xico,   A.P. 70-542, CDMX 04510, M\'{e}xico.}
\affiliation{$^2$Department of Physics, New York University,  726 Broadway, New York, NY10003, USA.}
\affiliation{$^3$Departamento de F\'{\i}sica, Universidad de Concepci\'on, Casilla, 160-C, Concepci\'on, Chile.}

\begin{abstract}
We construct higher-derivative gravity theories in three dimensions that admit holographic $c$-theorems and exhibit a unique maximally symmetric vacuum, at arbitrary order $n$ in the curvature. We show that these theories exhibit special properties, the most salient ones being the decoupling of ghost modes around Anti-de Sitter (AdS) space, the enhancement of symmetries at linearized level, and the existence of a one-parameter generalization of the Ba\~nados-Teitelboim-Zanelli (BTZ) black hole that, while being asymptotically AdS, is not of constant curvature but rather exhibits a curvature singularity. For such black holes, we provide a holographic derivation of their thermodynamics. This gives a microscopic picture of black hole thermodynamics for non-supersymmetric solutions, of non-constant curvature in higher-derivative theories of arbitrary order in the curvature.    
\end{abstract}
\maketitle

\section{Introduction}

Three-dimensional massive gravity \cite{Deser:1981wh, Bergshoeff:2009hq} has shown to be an excellent setup to investigate several aspects of holography and, in particular, to investigate to what extent the AdS/CFT correspondence can be generalized. Not only does it provide interesting backgrounds in which we can investigate the scope of the holographic paradigm beyond the Anti-de Sitter (AdS) space -- such as warped-AdS spaces \cite{Anninos:2008fx, Detournay:2012pc, Clement:2009gq, Donnay:2015iia, Donnay:2015vrb}, spaces with anisotropic scale invariance \cite{Ayon-Beato:2009cgh}, and asymptotically (A)dS black holes with non-constant curvature \cite{Oliva:2009ip, Bergshoeff:2009aq}--, but it also provides examples of different set of boundary conditions in AdS$_3$ itself, yielding a rich variety of relaxed falling off conditions near the boundary \cite{Grumiller:2008qz, Giribet:2011vv, Garbarz:2008qn}. The study of the latter has given rise to interesting discussions on how sensitive the properties of the dual conformal field theory (CFT) are to the prescription of specific boundary conditions. It has been shown that key aspects of the boundary CFT$_2$, such as chirality or unitarity, get affected by the relaxation of the standard Brown-Henneaux AdS$_3$ asymptotic conditions; see \cite{Li:2008dq, Maloney:2009ck, Giribet:2008bw, Carlip:2008eq, Grumiller:2008qz, Grumiller:2008es, Grumiller:2009sn}. For example, in \cite{Chernicoff:2024dll} we studied weakened AdS$_3$ boundary conditions in higher-curvature theories that yield logarithmic terms that are induced by the backreaction of quantum fluctuations in AdS$_3$. Here, we will consider a different type of relaxed AdS$_3$ asymptotics that are compatible with the same higher-curvature theories, which have a nice holographic interpretation in terms of $c$-theorems. In section \ref{sec:II}, we define the family of such theories, which can be regarded as higher-order generalization of the parity even New Massive Gravity (NMG) \cite{Bergshoeff:2009hq, Bergshoeff:2009aq}. Demanding the existence of a holographic $c$-theorem, we write down higher-order Lagrangian densities with terms of arbitrary order $n$ in the curvature. This generalizes the results in the literature about holographic $c$-teorems in three dimensions, cf. \cite{Sinha:2010ai, Paulos:2010ke}. When the additional requirement of the theory having a unique maximally symmetric ansatz is considered, the theories exhibit an additional conformal symmetry at the linearized level. This symmetry is associated to the existence of a new family of static black holes that generalize the Ba\~nados-Teitelboim-Zanelli (BTZ) solution. In section \ref{sec:III}, we check that the linearized equations of  motion of these theories are traceless, implying that, at this level, there is an extra conformal symmetry. In section \ref{sec:IV}, we study these black holes, which have interesting properties, such as the presence of a curvature singularity at the origin. We compute the conserved charges and the thermodynamic variables of these black holes, and we also reproduce their thermodynamics from the holographic point of view. We conclude in section \ref{sec:V} with some remarks.

\section{Theories satisfying a $c$-theorem with a single vacuum}\label{sec:II}

The set of three-dimensional higher-curvature gravity theories satisfying the existence of a holographic $c$-theorem \cite{Zamolodchikov:1986gt, Casini:2006es} and the set of similar theories that exhibit a single maximally symmetric vacuum were classified in \cite{Bueno:2022lhf}; see also \cite{Sinha:2010ai, Paulos:2010ke,Bergshoeff:2021tbz,Bueno:2022log}. Here, we are going to study theories that belong to the intersection of both sets. For the sake of clarity, we start reviewing the derivation at quadratic and cubic order, and later, we will explain the general construction to order $n$ in the curvature.

We say that a given three-dimensional gravity theory of order $n$,\footnote{We remember that in three dimensions, the Planck length can be written in terms of the gravitational constant as $\lp=8\pi G_{\text{N}}$.}
\begin{equation}
    I=\frac{1}{2\lp}\int\diff^3x\sqrt{-g}\Big( \mathcal{L}+\mathcal{L}_{\text{matt}}\Big) \, ,
\end{equation}
satisfies a simple holographic $c$-theorem if its field equations are at most of second order when evaluated on the following ansatz:
\begin{equation}\label{eq:intAdS}
\diff s^2=\diff r^2+\text{e}^{2A(r)}\left(-\diff t^2+\diff x^2\right),
\end{equation}
which interpolates between two asymptotically AdS${}_3$ regions; see \cite{Girardello:1998pd,Freedman:1999gp}. The field equations involve a suitable stress-energy tensor $T_{ab}$ associated to the matter Lagrangian $\mathcal{L}_{\text{matt}}$. $\mathcal{L}$ is the gravitational Lagrangian, which depends on the metric $g_{ab}$ and the Riemann tensor $R{}^a{}_{bcd}$. The two AdS$_3$ asymptotic regions represent the infrarred (IR) and the ultraviolet (UV) fixed points in the dual description, and, as usual, the interpolation itself is understood as the holographic realization of the renormalization group (RG) flow.

For theories of this sort, we can construct a quantity $c(r)$ which decreases monotonically along the RG flow and coincides with the Virasoro central charges of the CFT, $c$, at the respective fixed point. 
Consequently, the function $c(r)$ whose derivative with respect to $r$ obeys
\begin{equation}
    c'(r)\geq0\, , \quad \forall r\, ,
\end{equation}
is a good candidate to be the $c$-function in the dual picture. More concretely, according to \cite{Myers:2010tj, Myers:2010xs,Freedman:1999gp} we can consider
\begin{equation}
    c'(r)=-\frac{\pi}{A'}\left(T_t{}^t-T_r{}^r\right)\, ,
\end{equation}
and using the null energy condition (NEC), we have $T_{ab}\xi^a\xi^b\geq0$, where $\xi_a$ is a null vector, $\xi_a\xi^a=0$. For a perfect fluid of density $\rho$ and pressure $p$, we have $T_{ab}\xi^a\xi^b=\left[(\rho+p)u_au_b+pg_{ab}\right]\xi^a\xi^b=(\rho+p)(u_a\xi^a)^2\geq0$. Using $u^a=(1,0,0)$, with $u_au^a=-1$, we have $T_t{}^t=-\rho$ and $T_r{}^r=p$. Hence, NEC implies
\begin{equation}\label{eq:NEC}
    T_t{}^t-T_r{}^r \leq 0\, .
\end{equation}
The explicit expression of the $c$-function can be obtained from the Wald-like formula \cite{Wald:1993nt,Myers:2010tj,Sinha:2010ai}
\begin{equation}
    c(r)=\frac{\pi}{2 A'}\frac{\partial \mathcal{L}}{\partial R^{tr}{}_{tr}}\, ,
\end{equation}
where the derivative of the Lagrangian density $\mathcal{L}$ is to be evaluated on the interpolating ansatz \eqref{eq:intAdS}. Below, we will see how this works in concrete examples of different orders.

\subsection{Quadratic order}

As said, we are interested in higher-curvature three-dimensional gravity theories that satisfy the following two conditions: $a$) obeying simple holographic $c$-theorems as described above, and $b$) exhibit a unique maximally symmetric vacuum. Let us start by considering the well-known example of quadratic theories.

In three dimensions the Weyl tensor identically vanishes and so the Riemann curvature tensor is totally determined by the Ricci tensor and the metric. Because of this, we can write the most general theory as a function of the Ricci tensors $R_{ab}$ and the Ricci scalar $R$. In particular, this means that, at second order in curvature, it suffices to consider only two quadratic curvature invariants, $R^2=R_a^aR_b^b$ and $\R_2=R_a^bR_b^a$; then, one can consider theories of the form $\mathcal{L}(g_{ab},R_{ab})$ without loss of generality. We can write the quadratic term of the Lagrangian, $\mathcal{F}^{(2)}$, as
\begin{equation}
\mathcal F^{(2)}=L^2\left(\alpha_1 R^2+\alpha_2\mathcal R_2\right)\, ,
\end{equation}
with $L$ being some length scale so that the coupling constants $\alpha_1$ and $\alpha_2$ are dimensionless. The field equations coming from $\mathcal{F}^{(2)}$, which we write $K^{(2)}_{ab}=\lp T_{ab}^{(2)}$, are
\begin{align}
\frac{K_{ab}^{(2)}}{L^2}&=2\alpha_1R_{ab}R-\frac{1}{2}g_{ab}\left(\alpha_1R^2+\alpha_2\R_2\right)+2\alpha_1\left(g_{ab}\dal-\nabla_a\nabla_b\right)R+2\alpha_2R_a^cR_{bc}\notag\\
&+\frac{\alpha_2}{2}g_{ab}\dal R+\alpha_2\dal R_{ab}-2\alpha_2\nabla_c\nabla_{(a}R_{b)}^c\, ,
\end{align}
where we use the notation $2A_{(a}B_{b)}=A_aB_b+A_bB_a$ for tensor product symmetrization. Evaluating in the interpolating AdS ansatz \eqref{eq:intAdS}, we easily see that the combination ${T^{(2)}}_t^t-{T^{(2)}}_r^r$ reads
\begin{equation}
    \frac{\lp}{L^2} \left({T^{(2)}}_t^t-{T^{(2)}}_r^r\right)=4(3\alpha_1+\alpha_2)(A')^2A''-(8\alpha_1+3\alpha_2)\left[4(A'')^2+2A'A^{(3)}+A^{(4)}\right]\, .
\end{equation}
From here, we see that the choice $\alpha_2=-\frac{8}{3}\alpha_1$ reduces the equation of motion to second order. The resulting theory is the Bergshoeff-Hohm-Townsend gravity, also known as {New Massive gravity} (NMG) \cite{Bergshoeff:2009aq}, with the couplings $\alpha_1$ and $\alpha_2$ being proportional to the inverse of the squared graviton mass. This choice is particularly suitable for a simple holographic $c$-theorem, as the NEC \eqref{eq:NEC} is satisfied provided that $A''<0$.

Let us mention that the existence of a simple holographic $c$-theorem can also be inferred without making use of the ansatz (\ref{eq:intAdS}). It simply follows from the fact that, for the choice $\alpha_2=-\frac{8}{3}\alpha_1$, the trace of the field equations, $K^{(2)}=g^{ab}K_{ab}^{(2)}$, is of second order, yielding
\begin{equation}\label{eq:Kab2}
\frac{K^{(2)}}{L^2}=\frac{1}{2}\left(\alpha_1R^2+\alpha_2\R_2\right)+\frac{1}{4}\left(\alpha_1+\frac{3}{8}\alpha_2\right)\dal R\, .
\end{equation}
This implies the absence of a scalar mode around the maximally symmetric vacuum, reducing by one the number of local degrees of freedom and rendering the theory dynamically healthy. This is precisely why NMG is an interesting massive model: it lacks a ghost mode around AdS.

Now, if we include the Einstein-Hilbert and the cosmological term in the action, the full theory up to $n=2$ reads\footnote{From now on, when we write $\mathcal{L}^{(n)}$ we include in the Lagrangian all curvature orders up to order $n$.}
\begin{equation}\label{eq:L2pre}
    \mathcal{L}^{(2)}=R+\frac{2}{L^2}+\alpha_1L^2\left(R^2-\frac{8}{3}\R_2\right)\, ,
\end{equation}
whose vacuum field equations are given by
\begin{equation}\label{eq:Eab2}
\E^{(2)}_{ab}=G_{ab}-\frac{1}{L^2}g_{ab}+K^{(2)}_{ab}=0\, ,
\end{equation}
where $G_{ab}=R_{ab}-\frac{1}{2}g_{ab}R$ is the Einstein tensor and $K_{ab}^{(2)}$ is given in \eqref{eq:Kab2}. The quadratic theory admits, in general, two maximally symmetric backgrounds (two vacua). In order for the theory to have a unique vacuum, we must fix the value of the constant $\alpha_1$. To do so, we consider the field equations \eqref{eq:Eab2} and evaluate them on Einstein metrics of the form\footnote{In turn, Einstein metrics also satisfy $\bar R=-\frac{6}{\Ls^2}$ and $\bar \R_2=\frac{24}{\Ls^4}$.}
\begin{equation}\label{einstein}
    \bar R_{ab}=\frac{2}{\Ls^2}\bar g_{ab}\, ,
\end{equation}
where $\Ls$ is the AdS radius (in contrast to length scale $L$, which gives the bare cosmological constant $-L^{-2}$). Notice that the Einstein condition \eqref{einstein}, in 2+1 dimensions imply that the spacetime is of constant Riemann curvature, namely $R^{ab}_{\ \ cd}=-\frac{1}{\Ls^2}\delta^{ab}_{cd}$. The result of such evaluation is a quadratic equation of the quotient $\chi_0={L^2}/{\Ls^2}$ relating the bare and effective cosmological constants through the quadratic equation
\begin{equation}
\bar{\mathcal{E}}_{ab}=0\iff 1-\chi_0-\frac{2\alpha_1}{3}\chi^2_0=0\, ,
\end{equation}
which in general has two solutions. This equation admits a unique solution (a unique vacuum) for $\chi_0=2$. This requirement fixes the second available coupling constant in \eqref{eq:L2pre} and defines the one-parameter quadratic Lagrangian density \cite{Oliva:2009ip}
\begin{equation}\label{eq:D2}
\mathcal{D}^{(2)}=R+\frac{2}{L^2}-L^2\left(\frac{3}{8}R^2-\R_2\right)\, .
\end{equation}
This theory, at this very special point of the parameter space, exhibits very interesting properties such as enhanced symmetry at linearized level. For $L^2<0$ this special point agrees with the partially massless point in dS$_3$.

\subsection{Cubic order}

Now, let us take a look at the cubic order: A Lagrangian density of cubic order is constructed in a similar way, by combining $R^3$ and the product $R\R_2$ together with three contracted Ricci tensors $\R_3=R_a^bR_b^cR_c^a$. That is
\begin{equation}
\mathcal F^{(3)}=L^4\left(\alpha_3 R^3+\alpha_4R \R_2+\alpha_5 \R_3\right)\, .\label{L16}
\end{equation}
In this case, the piece of the field equations coming from (\ref{L16}) reads
\begin{align}\label{eq:Kab3}
    \frac{K_{ab}^{(3)}}{L^4}&=R_{ab}\left(3\alpha_3 R^2+\alpha_4\R_2\right)-\frac{1}{2}g_{ab}\left(\alpha_3 R^3+\alpha_4R \R_2+\alpha_5 \R_3\right)+2\alpha_4RR_a^cR_{bc}\notag\\
    &+\left(g_{ab}\dal-\nabla_a\nabla_b\right)\left(3\alpha_3R^2+\alpha_4\R_2\right)+\dal\left(\alpha_4R R_{ab}+\frac{3\alpha_5}{2}R_a^cR_{bc}\right)+3\alpha_5 R_a^cR_b^dR_{cd}\notag\\
    &+g_{ab}\nabla_c\nabla_d\left(\alpha_4RR^{cd}+\frac{3\alpha_5}{2}R^{cf}R_f^d\right)-2\nabla_c\nabla_{(a}\left(\alpha_4R R_{b)}^c+\frac{3\alpha_3}{2}R_{b)}^dR_d^c\right)\,
\end{align}
which, evaluated on the interpolating AdS ansatz \eqref{eq:intAdS}, yields
{\small \begin{align}
    \frac{\lp}{L^4} \left({T^{(3)}}_t^t-{T^{(2)}}_r^r\right)&=-36 (A')^4 A'' (9 \alpha_3+3 \alpha_4+\alpha_5)+\frac{1}{3} (72 \alpha_3+25 \alpha_4+9 \alpha_5) \left[5
   \left(A^{(3)}\right)^2+16 \left(A''\right)^3\right.\notag\\
    &\left.+5 A^{(4)} A''+12 A^{(3)} A'(\rho
   )^3+6 \left(A'\right)^2 \left(A^{(4)}+8 \left(A''\right)^2\right)+34 A^{(3)} A'A''\right]
\notag\\
&-\frac{1}{3} (72 \alpha_3+17 \alpha_4)
   \left[\left(A^{(3)}\right)^2+2 \left(A''\right)^3+A^{(4)} A''+2 A^{(3)} A'(\rho
   ) A''\right]
\end{align}}
We see that fixing two of the three coupling constants as $\alpha_4=-\frac{72\alpha_3}{17}$ and $\alpha_5=\frac{64\alpha_3}{17}$, we again have a theory satisfying a simple holographic $c$-theorem. As before, one could be tempted to derive the same constraints between $\alpha_3$, $\alpha_4$ and $\alpha_5$ by demanding the vanishing of higher-order terms in the trace of the field equations. However, at third order the constraints are much more involved than at second order.

We can now add the cubic theory $\mathcal F^{(3)}$ to the quadratic theory \eqref{eq:L2pre}, which also satisfies the same requirement. This yields the full theory
\begin{equation}
\mathcal{L}^{(3)}=R+\frac{2}{L^2}+\alpha_1L^2\left(R^2-\frac{8}{3}\R_2\right)+\alpha_3L^4\left(R^3-\frac{72}{17}R \R_2+\frac{64}{17} \R_3\right)\,.
\end{equation}
In total, the $c$-theorem fixes three coupling constants --one out of two at quadratic order and two out of three at cubic one-- and leaves two of them undetermined; we choose the latter to be $\alpha_1$ and $\alpha_3$. Now, proceeding in a similar way as in the previous subsection, we demand the existence of a unique vacuum: this amounts to evaluate the field equations
\begin{equation}
\E^{(3)}_{ab}=G_{ab}-\frac{1}{L^2}g_{ab}+K^{(2)}_{ab}+K^{(3)}_{ab}=0\, ,
\end{equation}
on the Einstein manifold ansatz $\bar R_{ab}=-\frac{2}{\Ls^2}g_{ab}$. Using the expression for $K^{(2)}_{ab}$ and $K^{(3)}_{ab}$ given in \eqref{eq:Kab2} and \eqref{eq:Kab3}, we find
\begin{equation}
   \bar{\E}^{(3)}_{ab}=0\iff 1-\chi_0-\frac{2\alpha_1}{3}\chi_0^2+\frac{12\alpha_3}{17}\chi_0^3=0\, ;
\end{equation}
and, from this, we see two conditions to get a single value $\chi_0=3$, namely $\alpha_1=-\frac{1}{2}$ and $\alpha_3=-\frac{17}{324}$. After imposing these constraints, we find the third order extension of theory \eqref{eq:D2}, whose Lagrangian density is
\begin{equation}\label{eq:D3}
    \mathcal D^{(3)}=R+\frac{2}{L^2}-\frac{L^2}{6}\left(3R^2-8\R_2\right)+\frac{L^4}{324} \left(17 R^3+72 R \R_2-64 \R_3\right)\, .
\end{equation}

As the quadratic theory defined by the Lagrangian density (\ref{eq:D2}), the cubic theory coming from (\ref{eq:D3}) has interesting features; among them, extra symmetries at linearized level and the existence of a rich variety of black hole solutions. We will discuss these features later.

\subsection{Arbitrary order $n$}

After reviewing the procedure to obtain higher-curvature theories that $a$) satisfy a simple holographic $c$-theorem and $b$) possess a single vacuum for the cases $n=2$ and $n=3$, we now move to arbitrary $n$. As mentioned before, the space of higher-curvature theories is substantially reduced in three dimensions due to the vanishing of the Weyl tensor. This allows for a classification of Lagrangians that is substantially simpler than in higher dimensions. All the independent theories are obtained from contractions of the Ricci tensor and the metric, $\mathcal L(g_{ab},R_{ab})$. Naively, when making the counting of independent densities, as we did for $n=2$ and $n=3$, one might expect that, at each order $n$, a new order-$n$ operator constructed from the Ricci tensor appears. However, it turns out that, for $n\geq 4$, one can remove such additional operators by using Schouten identities $\delta_{b_1\ldots b_n}^{a_1\ldots a_n} R_{a_1}^{b_1}\cdots R_{a_n}^{b_n}\equiv 0$, $n>3$. As a consequence, the most general gravity theory in three dimensions involving the Ricci tensor and its contractions at order $n$ can be written as \cite{Paulos:2010ke}
\begin{equation}
I=\frac{1}{2\ell_\text{P}}\int\diff^3x\sqrt{|g|}\mathcal{L}^{(n)}\, , \quad \mathcal{L}^{(n)}=R+\frac{2}{L^2}+\sum_{n}\mathcal{F}^{(n)}(R,\mathcal R_2,\mathcal R_3)\, ,
\end{equation}
where $\mathcal R_2=R^a_bR^b_a$ and $\mathcal R_3=R^a_bR^b_cR^c_a$. Alternatively, we can express the Lagrangian density in terms of contractions of the traceless Ricci tensor $\tilde R_{ab}=R_{ab}-\frac{1}{3}g_{ab}R$. Defining $\mathcal S_2=\tilde R^a_b\tilde R^b_a$ and $\mathcal S_3=\tilde  R^a_b\tilde  R^b_c \tilde R^c_a$, we will write the most general Lagrangian density as follows
\begin{equation}\label{eq:S3DS}
\mathcal L^{(n)}=R+\frac{2}{L^2}+\sum_{n}\mathcal{G}^{(n)}(R,\mathcal S_2,\mathcal S_3)\, ;
\end{equation}
this basis turns out to be more convenient to derive and present the results. We focus on higher-curvature theories that admit a polynomial expression, i.e.
\begin{align}
    \mathcal{F}^{(n)}(R,\mathcal R_2,\mathcal R_3)&=\sum_{i+2j+3k=n}\alpha_{ijk}L^{
2(i+2j+3k-1)}R^i\mathcal{R}_2^j\mathcal{R}_3^k\, ,\\
\mathcal{G}^{(n)}(R,\mathcal S_2,\mathcal S_3)&=\sum_{i+2j+3k=n}\beta_{ijk}L^{
2(i+2j+3k-1)}R^i\mathcal S_2^j\mathcal S_3^k\, ,\label{eq:Gn}
\end{align}
where $\alpha_{ijk}$ and $\beta_{ijk}$ represent the dimensionless coupling constants for each order-$n$ term.

The higher-curvature terms of order $n$ satisfying a simple holographic $c$-theorem are given by special values of $\alpha_{ijk}$ and $\beta_{ijk}$; we denote those Lagrangian densities $\mathcal{F}^{(n)}_{c\text{-theorem}}$, which were previously identified in \cite{Bueno:2022lhf}. They obey
\begin{equation}\label{eq:Lc}
\mathcal F_{c\text{-theorem}}^{(n)}=\gamma_n\mathcal{C}^{(n)}+\Omega^{(6)}\cdot\mathcal F_{\text{general}}^{(n-6)}\, ,
\end{equation}
where\footnote{Note that we do not use the expression for $\mathcal C^{(n)}$ given in \cite{Bueno:2022lhf}. Instead, we use the three-dimensional case of the Lagrangian with the same properties presented in \cite{Moreno:2023arp}, which is a simpler expression.}
\begin{equation}\label{eq:Cn}
\mathcal C^{(n)}= R^n-\sum_{p=1}^{\lfloor n/2\rfloor}24^p\binom{n}{2p}R^{n-2p}\mathcal S_2^p+\sum_{p=1}^{\lfloor (n-1)/2\rfloor}24^{p+1}p\binom{n}{2p+1}R^{n-2p-1}\mathcal S_2^{p-1}\mathcal S_3
\end{equation}
and $\Omega^{(6)}=6\mathcal S_3^2-\mathcal S_2^3$ is a density that vanishes identically when evaluated on the ansatz \eqref{eq:intAdS}. As shown in \cite{Bueno:2022lhf}, $\mathcal{C}^{(n)}$ is the only non-trivial density satisfying a holographic $c$-theorem. Notice that, as we checked in the quadratic and the cubic cases, imposing second-order equations for the ansatz \eqref{eq:intAdS} for a set of order-$n$ densities yield $n - 1$ conditions, leaving only one free constant: $\gamma_n$.

Now, let us turn our attention to the second requirement: the theory must possess a single vacuum. This demands additional $n-1$ constraints to the values of $\gamma_n$, with $\gamma_0$ and $\gamma_1$ being the bared cosmological constant and the bared Planck mass, respectively. 

The full field equations in vacuum for the general Lagrangian density \eqref{eq:S3DS} are given by
\begin{equation}
\mathcal{E}_{ab}^{(n)}=G_{ab}-g_{ab}\frac{1}{L^2}+\sum_nK^{(n)}_{ab}=0\, 
\end{equation}
where the piece corresponding to the order-$n$ invariant $\mathcal{G}^{(n)}$ reads 
\begin{align}
K_{ab}^{(n)}=&-\frac{1}{2}g_{a b}\mathcal{G}^{(n)}+2 \mathcal{G}_{\mathcal{S}_2}^{(n)}  \tilde{R}_{a}^{c}\tilde{R}_{c b}+ 3 \mathcal{G}_{\mathcal{S}_3}^{(n)}\tilde{R}_{a}^{c}\tilde{R}_{c d}\tilde{R}_{b}^{d}+g_{a b}\nabla_{c}\nabla_{d}\left(\mathcal{G}_{\mathcal{S}_2}^{(n)}\tilde R^{c d}+\frac{3}{2}\mathcal{G}_{\mathcal{S}_3}^{(n)}\tilde R^{c f}\tilde{R}_{f}^{d}\right)\notag\\ 
&+\left(g_{a b}\dal - \nabla_{a}\nabla_{b}+\tilde{R}_{a b}+\frac{1}{3}g_{a b}R\right) \left(\mathcal{G}_R^{(n)}-\mathcal{G}_{\mathcal{S}_3}^{(n)}\mathcal{S}_2\right)-2\nabla_{c}\nabla_{(a}\left(\tilde{R}_{b)}^{c}\mathcal{G}_{\mathcal{S}_2}^{(n)}+\frac{3}{2}\tilde{R}_{b)}^{d}\tilde{R}_{d}^{c}\mathcal{G}_{\mathcal{S}_3}^{(n)}\right)\notag\\
&+\left(\dal+\frac{2}{3}R\right) \left( \mathcal{G}_{\mathcal{S}_2}^{(n)}\tilde{R}_{a b}+\frac{3}{2}\mathcal{G}_{\mathcal{S}_3}^{(n)}\tilde{R}_{a}^{c}\tilde{R}_{c b}\right)=0 \, ,\label{eq:EOM3}
\end{align}
and $\mathcal{G}_{X}^{(n)}$ represents the partial derivative of the $\mathcal{G}^{(n)}$ with respect to the invariant $X$, i.e. $\mathcal{G}_X= \partial \mathcal{G}/\partial X$.

Now, consider AdS$_3$ spacetime: then, the three densities in our class of Lagrangian take the values\footnote{These expressions make manifest the convenience of using the basis $(R,\mathcal S_2, \mathcal S_3)$ instead of $(R,\mathcal R_2, \mathcal R_3)$, as $\bar{\mathcal{R}}_2=\frac{12}{\Ls^4}$, $\bar{\mathcal{R}}_3=\frac{24}{\Ls^6}$, making the general analysis using the latter more involved.}
\begin{equation}
\bar R=-\frac{6}{L^2_\star}\,, \quad \bar{\mathcal{S}}_2=0\,, \quad \bar{\mathcal{S}}_3=0\,.
\end{equation}
Substituting this into the full field equations $\bar{\mathcal{E}}_{ab}^{(n)}=0$, we obtain the simple relation
\begin{equation}\label{vacuS}
\frac{2}{ L^2}-\frac{2}{L_{\star}^2}+\sum_n\left(\bar{\mathcal{G}}^{(n)}+\frac{4}{\Ls^2}\bar{\mathcal{G}}_R^{(n)}\right)n=0\,.
\end{equation}
If we consider a Lagrangian expressed in terms of a polynomial as in \eqref{eq:Gn}; then, equation \eqref{vacuS} reduces to the characteristic polynomial
\begin{equation}
1-\chi_{0}+\sum_{n} b_n  \chi_{0}^n=0\,,
\end{equation}
where $b_n= (-1)^n 6^{n-1} (3-2n) \beta_{n00}$. Notice that, as anticipated above, the terms involving $\mathcal S_2$ and $\mathcal S_3$ do not contribute to the characteristic polynomial and so do not modify $\chi_0$.

A single-vacuum theory has order-$n$ degeneration of the solution to the characteristic polynomial. For them, the value of $\chi_0$ is given by
\begin{equation}
\chi_0=n\,,
\end{equation}
which generalizes what we have obtained for $n=2$ and $n=3$. Based on this observation, it was argued in \cite{Bueno:2022lhf} that a Lagrangian density of the form
\begin{equation}\label{eq:singv}
\mathcal{L}^{(n)}_{\text{single vac.}}=\frac{2}{L^2}+R+\sum_{i=2}^n \binom{n}{i}\frac{L^{2(i-1)}}{n^i 6^{i-1}(3-2i)}R^i + \mathcal{S}_2 h_2(R,\mathcal{S}_2,\mathcal{S}_3)+ \mathcal{S}_3 h_3(R,\mathcal{S}_2,\mathcal{S}_3)\, ,
\end{equation}
where $h_2$ and $h_3$ are any analytic functions of their arguments, possesses a single AdS$_3$ vacuum. Notice that, in these theories, the coupling constant of the order-$n$ curvature term scales as $\sim L^{2n-2}$, with the bared cosmological constant being $\sim L^{-2}$. This implies that the theory is strongly coupled, with the UV and the IR regimes being related. This is precisely the reason why, while higher-derivative terms are generically expected to yield short-distance modifications to general relativity, for these theories we still expect long-range modifications. This is what we will actually find: slow-decaying modes will alter the standard asymptotically AdS$_3$ behavior in an interesting manner. 

With this in mind, we can combine expressions \eqref{eq:Cn} and \eqref{eq:singv} to find a theory satisfying both requirements: $a$) possessing a holographic $c$-theorem and $b$) a single vacuum. The single vacuum requirement imposes a particular value of the coupling constant for the term $R^n$, however, it leaves unspecified what $h_2$ and $h_3$ should be, as they do not modify the characteristic polynomial. As a consequence, we can include the same coupling constant for the additional two terms appearing in \eqref{eq:Cn}, so that we do not spoil the relative factors that guarantee the existence of a holographic $c$-theorem at order $n$. In turn, we obtain the following Lagrangian
\begin{align}\label{eq:Lagchteosing}
\mathcal{D}^{(n)}=\frac{2}{L^2}+R&+\sum_{i=2}^n\binom{n}{i}\frac{L^{2(i-1)}}{n^i6^{i-1}(3-2i)}\left[R^i-\sum_{p=1}^{\lfloor i/2\rfloor}24^p(2p-1)\binom{i}{2p}R^{i-2p}\mathcal S_2^p\right.\notag\\
&\left.+\sum_{p=1}^{\lfloor (i-1)/2\rfloor}24^{p+1}p\binom{i}{2p+1}R^{i-2p-1}\mathcal S_2^{p-1}\mathcal S_3\right]\, ,\quad n\geq 1\, ,
\end{align}
which, of course, reproduces the quadratic and cubic theories $\mathcal D^{(2)}$ and $\mathcal{D}^{(3)}$, presented in \eqref{eq:D2} and \eqref{eq:D3} respectively, although in the $(R,\mathcal S_2, \mathcal S_3)$ basis, namely
\begin{align}
\mathcal{D}^{(2)}&=\frac{2}{L^2}+R-\frac{L^2}{24}\left(R^2-24\mathcal S_2\right)\, .\\
\mathcal{D}^{(3)}&=\frac{2}{L^2}+R-\frac{L^2}{18}\left(R^2-24\mathcal S_2\right)-\frac{L^4}{2916}\left(R^3-72R\mathcal S_2+576\mathcal S_3\right)\, .
\end{align}

The expressions for higher $n$ densities, $\mathcal{D}^{(n)}$, are more complicated. To illustrate that, let us show some examples up to order $n=6$
\begin{align}
\mathcal{D}^{(4)}&=\frac{2}{L^2}+R-\frac{L^2}{16}\left(R^2-24\mathcal S_2\right)-\frac{L^4}{1728}\left(R^3-72R\mathcal S_2+576\mathcal S_3\right)\, ,\\
&+\frac{L^6}{276480}\left(R^4 - 144 R^2 \mathcal S_2 - 1728 \mathcal S_2^2 + 2304 R \mathcal S_3\right),\notag\\
\mathcal{D}^{(5)}&=\frac{2}{L^2}+R-\frac{L^2}{15}\left(R^2-24\mathcal S_2\right)-\frac{L^4}{1350}\left(R^3-72R\mathcal S_2+576\mathcal S_3\right)\\
&+\frac{L^6}{135000}\left(R^4 - 144 R^2 \mathcal S_2 - 1728 \mathcal S_2^2 + 2304 R \mathcal S_3\right)\notag\\
&+\frac{L^8}{28350000}\left(R^5 - 240 R^3 \mathcal S_2 - 8640 R \mathcal S_2^2 + 5760 R^2 \mathcal S_3 + 27648 \mathcal S_2 \mathcal S_3\right)\, ,\notag\\
\mathcal{D}^{(6)}&=\frac{2}{L^2}+R-\frac{5L^2}{72}\left(R^2-24\mathcal S_2\right)-\frac{5L^4}{5832}\left(R^3-72R\mathcal S_2+576\mathcal S_3\right)\\
&+\frac{L^6}{93312}\left(R^4 - 144 R^2 \mathcal S_2 - 1728 \mathcal S_2^2 + 2304 R \mathcal S_3\right)\notag\\
&+\frac{L^8}{11757312}\left(R^5 - 240 R^3 \mathcal S_2 - 8640 R \mathcal S_2^2 + 5760 R^2 \mathcal S_3 + 27648 \mathcal S_2 \mathcal S_3\right)\notag\\
&+\frac{L^{10}}{3265173504}\left(R^6 - 360 R^4 \mathcal S_2 - 25920 R^2 \mathcal S_2^2 - 69120 \mathcal S_2^3 + 11520 R^3 \mathcal S_3 + 165888 R \mathcal S_2 \mathcal S_3\right)\notag\, .
\end{align}


The statement that, at each order $n$, there is a unique theory that satisfy at the same time $a$) having simple holographic $c$-theorem and $b$) having a unique maximally symmetric vacuum, has to be understood on-shell\footnote{We use the same definition of equivalent densities as the one introduced in \cite{Bueno:2022res}.}. In fact, we have the freedom of adding to $\mathcal{D}^{(n)}$ Lagrangian densities that do not alter the properties $a$) and $b$). An example of this is the deformed Lagrangian density ${\mathcal{D}^{(6)}}'=\mathcal{D}^{(6)}+\Omega^{(6)}$, which fulfill the same requirements. Still, ${\mathcal{D}^{(6)}}'$ and $\mathcal{D}^{(6)}$ are equivalent in what regards to our requirements, so that we can talk about an equivalence class of theories.

\section{Linearized field equations}\label{sec:III}

In this section we show that all the theories defined by the Lagrangian densities $\mathcal{D}^{(n)}$, given in \eqref{eq:Lagchteosing}, satisfy that the trace of the field equations vanish at linearized level, and that this is associated to the existence of a new gauge symmetry. To show this explicitly, recall that the field equations at leading order in perturbation theory for any three-dimensional model expressed in terms of \eqref{eq:S3DS} read \cite{Bueno:2022lhf}
\begin{equation}\label{line}
\frac{1}{4}{\mathcal{E}^{(n)}_{ab}}\lnr=\left[e+c\left(\bar{\dal}+\frac{2}{L_{\star}^2}\right)\right]G\lnr_{ab}+(2b+c)\left(\bar{g}_{ab}\bar{\dal}-\bar{\nabla}_a\bar{\nabla}_b\right)R\lnr-\frac{1}{L_{\star}^2}\left(4b+c\right)\bar{g}_{ab}R\lnr=\frac{\ell_{\text P}}{4}T_{ab}\lnr\, ,\small
\end{equation}
where $T_{ab}\lnr$ is the linearized stress-energy tensor and where the linearized Einstein tensor, Ricci tensor, and Ricci scalar are given by
 \begin{align}
G\lnr_{ab}&=R\lnr_{ab}-\frac{1}{2}\bar{g}_{ab}R\lnr+\frac{2}{L_{\star}^2} h_{ab}\, ,\\
R\lnr_{ab}&=\bar{\nabla}_{\left(a\right|}\bar{\nabla}_{c}h\indices{^c_{\left|b\right)}}-\frac{1}{2}\bar{\dal}h_{ab}-\frac{1}{2}\bar{\nabla}_a\bar{\nabla}_b h-\frac{3}{L_{\star}^2} h_{ab}+\frac{1}{L_{\star}^2} h \bar{g}_{ab}\, ,\\
R\lnr &=\bar{\nabla}^a\bar{\nabla}^b h_{ab}-\bar{\dal} h+\frac{2}{L_{\star}^2} h\, ,
\end{align}
respectively. The parameters $e$, $c$ and $b$ in (\ref{line}) are to be interpreted as the effective Planck length $\lp^{\text{eff}}$ and the masses (squared) of gravitational modes, which we denote $m_g^2$ and $m_s^2$, as
\begin{equation}\label{phypara}
\ell_{\text P}^{\text{eff}}=\frac{1}{4e}\, , \quad m_g^2=-\frac{e}{c}\, , \quad m_s^2=\frac{e+\frac{8}{L_{\star}^2}(3b+c)}{3c+8b}\, .
\end{equation}
They are obtained from the relations
\begin{equation}\label{GGL}
e=\frac{1}{4\ell_{\text P}} [1+\bar{\mathcal{G}}_R]\, , \quad b=\frac{1}{4\ell_{\text P}} \left[\frac{1}{2}\bar{\mathcal{G}}_{R,R}-\frac{1}{3} \bar{\mathcal{G}}_{\mathcal{S}_2} \right]\, , \quad c=\frac{1}{4\ell_{\text P}} \bar{\mathcal{G}}_{\mathcal{S}_2}\, ,
\end{equation}
where we used again the notation $\mathcal{G}_X= \partial \mathcal{G}/\partial X$, $\mathcal{G}_{X,X}= \partial^2 \mathcal{G}/\partial X^2$.

Taking this into account, we can compute the specific values of the parameters $e$, $b$ and $c$ parameters for the theory $\eqref{eq:Lagchteosing}$. This gives
\begin{equation}\label{eq:ecb}
e=\frac{1}{4\lp}\left[1+\sum_{i=2}^n\binom{n}{i}\frac{(-1)^{i-2}i}{n(3-2i)}\right]=\frac{1}{4\lp}\frac{\sqrt{\pi}\,\Gamma(n)}{\Gamma(n-1/2)}\,,\quad c=-\Ls^2e\,,\quad b=-\frac{3}{8}c\,.
\end{equation}
In turn, this implies that the effective Planck length is given by
\begin{equation}\label{eq:lP}
\ell_{\text P}^{\rm eff}=\lp\frac{\Gamma\left(n-\frac{1}{2}\right)}{\sqrt{\pi}\Gamma(n)}\, ,
\end{equation}
while the other physical parameters read
\begin{equation}\label{eq:params}
m_g^2=-\frac{1}{\Ls^2}=-\frac{n}{L^2}\, , \quad m_s^2\rightarrow\infty\, .
\end{equation}
The latter shows that the ghost scalar modes decouples, generalizing what occurs in NMG. 

Without imposing any particular value for $e$, $b$ and $c$ yet, the trace of the linearized field equations reads
\begin{equation}\label{GG48}
\bar{g}^{ab}\mathcal{E}\lnr_{ab}=(8b+3c)\bar{\dal}R\lnr-\left(\frac{24b}{L_\star^2}+\frac{8c}{L_\star^2}+e\right)R\lnr\,.
\end{equation}
We see that the term proportional to $\dal R\lnr$ vanishes as a consequence of imposing a $c$-theorem \eqref{eq:ecb}. Using the other relations, which implement the single-vacuum condition, we see that the coefficient of $R\lnr$ in the second term of (\ref{GG48}) vanishes too, and in that case we obtain
\begin{equation}
\E\lnr=0\,.
\end{equation}
This means that the theories we are considering have an extra conformal symmetry at linearized level around constant curvature backgrounds, which corresponds to a Weyl symmetry transformation. This symmetry is associated to the existence of a family of conformally flat solutions that generalize the BTZ black hole \cite{Gabadadze:2012xv}. We will study these solutions in the next section.  

\section{Hairy black holes}\label{sec:IV}

\subsection{The solution}

We now proceed to prove that our theories \eqref{eq:Lagchteosing} admit a one-parameter family of exact solutions that generalize the BTZ black hole. The metric is 
\begin{equation}\label{eq:hbh}
\diff s^2_{\text {hbh}}=-f(r)\diff t^2+\frac{\diff r^2}{f(r)}+r^2\diff \phi^2\,, \quad f(r)=\frac{r^2}{\Ls^2}+b r-\lambda\,,
\end{equation}
which, indeed, reduces to the metric of the static BTZ geometry when $b=0$. A stationary generalization including an angular momentum parameter $a$ also exists \cite{Oliva:2009ip}. The parameter $b$ is an additional integration constant that can be regarded as a gravitational hair, so that we refer to (\ref{eq:hbh}) as the hairy black hole (that is what the subindex $\text {hbh}$ in (\ref{eq:hbh}) stands for). When the hair parameter $b$ is different from zero, the solution exhibits a curvature singularity at the origin; the expansion of the Ricci scalar around $r=0$ is $R=-\frac{2b}{r}+\mathcal{O}(r^0)$. The solution has a rich causal structure, depending on the interplay between the parameters $b$, $\Ls$ and $\lambda$. The black holes have a single horizon $r_+$ unless $b<0$ and $-b^2\Ls^2<4\lambda<0$; in the latter cases, it may have two horizons, $r_+$ and $r_-$, given by
\begin{equation}
r_\pm=\frac{\Ls}{2}\left(\pm\sqrt{b^2\Ls^2+4\lambda}-b\Ls\right)\, , \quad r_+-r_-=\Ls\sqrt{b^2\Ls^2+4\lambda}\,.
\end{equation}
Notice that when $-b^2\Ls^2=4\lambda$ the two horizons coincide and the solution becomes extremal.

Let us now compute the quantities $\mathcal G$, $\mathcal G_R$, $\mathcal G_{\mathcal S_2}$ and $\mathcal G_{\mathcal S_3}$ for this solution: The first two take simple expressions, namely
\begin{align}
    \mathcal G\big|_{\diff s^2_{\text {hbh}}}=&\frac{2 b}{r}+\frac{2
   }{\Ls^2}\left(3-\frac{1}{n}\right)-\frac{\sqrt{\pi } \Gamma (n)}{\Gamma
   \left(n-\frac{1}{2}\right)}\left(\frac{2 b}{r}+\frac{4}{\Ls^2}\right)\,,\\
   \mathcal G_{R}\big|_{\diff s^2_{\text {hbh}}}=&\frac{\sqrt{\pi } \Gamma (n)}{\Gamma
   \left(n-\frac{1}{2}\right)}\left(\frac{b \Ls^2}{6 r}+1\right)-1\, . \label{eq:GRbar}
\end{align}
$\mathcal G_{\mathcal S_2}$ and $\mathcal G_{\mathcal S_3}$ take a more involved form, namely
\begin{align}
    \mathcal G_{\mathcal S_2}\big|_{\diff s^2_{\text {hbh}}}=&\frac{\sqrt{\pi } \Gamma (n)}{\Gamma
   \left(n-\frac{1}{2}\right)}\left[\Ls^2+\frac{r}{b}\left(9-\sqrt{9+\frac{6b\Ls^2}{r}}\right)+\frac{r^2}{b^2\Ls^2}\left(3-\sqrt{9+\frac{6b\Ls^2}{r}}\right)\right.\\
   &\left.-\left(-\frac{b}{r}\right)^{n-1}\frac{6^{2-n}\Ls^{2n}\Gamma(2n-2)}{(2b\Ls^2+3r)\Gamma(n-1)\Gamma(n+1)}{}_2\tilde{F}_1\left(1,n-\frac{1}{2},n+2;-\frac{2b\Ls^2}{3r}\right) \right]\, ,\notag\\
   \mathcal G_{\mathcal S_3}\big|_{\diff s^2_{\text {hbh}}}=&\frac{\sqrt{\pi } \Gamma (n)}{\Gamma
   \left(n-\frac{1}{2}\right)}\left\{\frac{4 r}{b^3 \Ls^2} \left[b \Ls^2 \left(9 r-\sqrt{6 b \Ls^2 r+9 r^2}\right)-6 \sqrt{6 b \Ls^2 r^3+9 r^4}+18 r^2\right]\right.\\
   &+\frac{n (n+1) \Ls^{2 n} \Gamma (2 n-1) }{2^{n-3} 3^{n-2}\Gamma (n+2)}\left(-\frac{b}{r}\right)^{n-2}\left[{}_2\tilde{F}_1\left(2,n-\frac{1}{2},n+2;-\frac{2b\Ls^2}{3r}\right)\right.\notag\\
   &+\left.\left.(n-2){}_2\tilde{F}_1\left(1,n-\frac{1}{2},n+2;-\frac{2b\Ls^2}{3r}\right)\right]\right\}\, ,\notag
\end{align}
where ${}_2\tilde{F}_1\left(a,b,c;z\right)={{}_2F_1\left(a,b,c;z\right)}/{\Gamma(c)}$ is the regularized hypergeometric function.

Having these expressions, we can plug them into the field equations \eqref{eq:EOM3} and verify that they are indeed satisfied. This proves the existence of the hairy black holes \eqref{eq:hbh} for the family of theories defined by the Lagrangian density \eqref{eq:Lagchteosing}. This is a generalization of the NMG hairy black hole solutions to all order in the curvature. 

Since the theories defined by \eqref{eq:Lagchteosing} obey the holographic $c$-theorem condition, it is natural to undertake a holographic analysis of the thermodynamics of the hairy black holes (\ref{eq:hbh}). This is more interesting than the case of BTZ solution because, on the one hand, solution (\ref{eq:hbh}) is not of constant curvature and, consequently, is not locally equivalent to AdS$_3$. On the other hand, the existence of an extra parameter makes the matching with the dual CFT quantities less evident.

\subsection{Central charge}

In preparation to the holographic computation of the black hole thermodynamics, let us first compute the central charge of the CFT that would be dual to the theories defined by (\ref{eq:Lagchteosing}). This amounts to use the formula
\begin{equation}\label{eq:centralc}
    c=\frac{4\pi\Ls}{\ell_{\text{P}}}g_{ab}\frac{\partial \mathcal {D}^{(n)}}{\partial R_{ab}} .
\end{equation}
Explicitly, we have
\begin{align}
    \frac{\partial \mathcal D_{(n)}}{\partial R_{ab}}=&g^{ab}+\sum_{i=2}^n\binom{n}{i}\frac{L^{2(i-1)}}{n^i6^{i-1}(3-2i)}\Biggr\{i R^{i-1}g^{ab}\notag\\
    &-\sum_{p=1}^{\lfloor i/2\rfloor}24^p(2p-1)\binom{i}{2p}\left[(i-2p)R^{i-2p-1}\mathcal S_2^p g^{ab}+2pR^{i-2p}\mathcal S_2^{p-1}\tilde R_c^a\tilde R^{bc}\right]\notag\\
&+\sum_{p=1}^{\lfloor (i-1)/2\rfloor}24^{p+1}p\binom{i}{2p+1}\left[(i-2p-1)R^{i-2p-2}\mathcal S_2^{p-1}\mathcal S_3g^{ab}\right.\notag\\
&\left.+(2p-1)R^{i-2p-1}\mathcal S_2^{p-2}\mathcal S_3\tilde R_c^a\tilde R^{bc}+R^{i-2p-1}\mathcal S_2^{p-1}\mathcal S_3\left(3\tilde R_c^a\tilde R^{bc}-\mathcal S_2 g^{ab}\right)\right]\Biggr\}\, .\label{eq:partialD}
\end{align}
After evaluating this expression on AdS${}_3$ (i.e. $b=0$ and $\lambda=-1$ in \eqref{eq:hbh}) and inserting the result back in \eqref{eq:centralc}, we obtain
\begin{equation}
    c=\frac{12\pi \Ls}{\ell_{\text{P}}}\frac{\sqrt{\pi} \Gamma(n)}{\Gamma\left(n-\frac{1}{2}\right)}=\frac{12\pi \Ls}{\ell_{\text{P}}^{\text{eff}}}\, ,
\end{equation}
where in the second equality we used the relation between $\ell_{\text{P}}^{\text{eff}}$ and $\ell_{\text{P}}$ given in \eqref{eq:lP}. For Einstein gravity ($n=1$) we recover the Brown-Henneaux central charge $c=3L/(2G_{\text N})$ \cite{Brown:1986nw}.

\subsection{Conserved charges and thermodynamics}
Let us now compute the thermodynamic variables and the conserved charges associated to the black hole solution (\ref{eq:hbh}). The Hawking temperature is
\begin{equation}
    T=\frac{f'(r_+)}{4\pi}=\frac{1}{4\pi \Ls}\sqrt{b^2\Ls^2+4\lambda}=\frac{r_+-r_-}{4\pi \Ls^2}\, ,
\end{equation}
which vanishes for the extremal solution $r_+=r_-$.

The mass of the solution can be easily computed by making use of the following trick: defining the variable ${r'}=r+\frac{b\Ls^2}{2}$ the metric function takes the form $f({r'})=\frac{{r'}^2}{\Ls^2}-\left(\lambda+\frac{b^2\Ls^2}{4}\right)$, which is the usual form of the BTZ metric but having shifted the mass parameter as $\lambda\mapsto \lambda+\frac{b^2\Ls^2}{4}$. Of course, this also changes the metric component $g_{\phi\phi}$; however, it does so in a subleading contribution $\mathcal{O}({r'})$ that does not affect the flux integral in the charge computation. Therefore, the mass of the hairy solution simply corresponds to the mass of the BTZ in the higher-curvature model, $M_{(b=0)}=\frac{\pi}{\lp^{\text{eff}}}\lambda$, but with a shift in $\lambda$. More precisely, we have the mass
\begin{equation}
    M=\frac{\pi}{4\lp^{\text{eff}}}\left(b^2\Ls^2+4\lambda\right)=\frac{\pi \left(r_+-r_-\right)^2}{4\lp^{\text{eff}} \Ls^2}\, .
\end{equation}
This result is confirmed by the explicit calculation of the mass using the Abbott-Deser-Tekin method (see \cite{Tekin:2023uks} and references therein), which we explicitly checked up to order $n=6$. 

Now we turn our attention to the Bekenstein-Hawking entropy $S$, which can be computed using the Wald entropy formula. This formula states that, for theories constructed with the Ricci tensor and its contractions, the following identity holds
\begin{equation}\label{eq:WaldE}
    S=-\frac{\pi}{\ell_{\text{P}}}\int_{\Sigma_h}\frac{\partial \mathcal L}{\partial R_{ab}}\epsilon_{ac}\epsilon_b{}^c \omega\, ,
\end{equation}
where $\omega=\sqrt{h}\, \diff \phi$ represents the volume 1-form of the induced metric $h_{ab}$ and $\epsilon_{ab}$ is a bi-normal vector to the space-like bifurcation surface $\Sigma_h$. As we consider the theories defined with the densities $\mathcal{D}^{(n)}$, we can use expression \eqref{eq:partialD} and evaluate it on the hairy black hole \eqref{eq:hbh}. We use that the bi-normal vector is $\epsilon_{ab}=-2\delta^t_{[a}\delta^r_{b]}$ with the normalization $\epsilon_{ab}\epsilon^{ab}=-2$. With all this, we obtain that the entropy is
\begin{equation}\label{BBHH}
    S=\frac{2\pi^2 \Ls}{\ell_\text{P}}\frac{\sqrt{\pi}\Gamma(n)}{\Gamma\left(n-\frac{1}{2}\right)}\sqrt{b^2\Ls^2+4\lambda}=\frac{2\pi^2(r_+-r_-)}{\ell_\text{P}^{\text{eff}}}\, ,
\end{equation}
which reduces to the standard expression for BTZ when $b=0$. This result for the black hole entropy, together with the Hawking temperature and the mass computed above, can be shown to satisfy the first law of black hole mechanics 
\begin{equation}
\diff M\, =\, T\, \diff S\, ,\quad  \text{with} \quad \diff = \diff \lambda\, \frac{\partial}{\, \partial \lambda |_{b}}+
\diff b\, \frac{\partial}{\, \partial b |_{\lambda}}\, ,
\end{equation}
i.e. with the functional variation being defined in terms of both $\lambda$ and $b$. Also, combining the results above, we can also compute the free energy
\begin{equation}
    F=M-TS=-\frac{\pi\left(r_+-r_-\right)^2}{2\lp^{\text{eff}}\Ls^2}\,,
\end{equation}
which does vanish for the extremal solutions.

\subsection{Holographic derivation}

Alternatively, we can compute the black hole entropy using holography. Concretely, we can resort to the relation that is expected to be satisfied by the central charge of the dual CFT, the Hawking temperature and the Bekenstein-Hawking entropy; namely
\begin{equation}
    S=\frac{2\pi^ 2\Ls c}{3} \, T\, ,\label{yestatambien}
\end{equation}
which, in fact, is found to match the result (\ref{BBHH}). This agreement is known to hold for the BTZ solution in arbitrary higher-curvature model, and the proof of that statement makes use of the fact that BTZ is locally equivalent to AdS$_3$ \cite{Saida:1999ec}. Here, in contrast, we are verifying this for a more general type of geometry, which are not of constant curvature but rather exhibit a curvature singularity at the origin.

In the case the black hole presents two horizons ($b<0$) we can use that $2\pi \left(r_+- r_-\right)=2\pi\Ls\sqrt{\Ls^2b^2+4\mu}=A_+-A_-$, to express the entropy as follows
\begin{equation}
    S=\frac{\pi(A_+-A_-)}{\ell_\text{P}^{\text{eff}}}\, ,
\end{equation}
where $A_{\pm }=2\pi r_{\pm }$ is the perimeter of the horizon at $r_{\pm }$. This form of writing the entropy is convenient as it permits a further check of the holographic expressions. In fact, for the static solution, the eigenvalues of the Virasoro generators $L_0+\bar{L}_0$ and $L_0-\bar{L}_0$ are given by
\begin{equation}
    \Delta+\bar{\Delta}={\Ls M}\, , \quad  \Delta-\bar{\Delta}=0\, ,
\end{equation}
respectively. This enables to check that the usual form of the Cardy formula holds; namely
\begin{equation}
    S=2\pi\sqrt{\frac{c\, \Delta}{6}}+2\pi\sqrt{\frac{c\, \bar{\Delta}}{6}}\, ;
    \end{equation}
this exactly matches (\ref{BBHH}) and (\ref{yestatambien}). A generalization of this calculation for the case of stationary black holes ($\Delta-\bar{\Delta}\neq 0$) is also doable, but the expressions in that case are much more cumbersome and we prefer to omit the details.

\section{Conclusions}\label{sec:V}

In this paper we have generalized results of references \cite{Oliva:2009ip, Bergshoeff:2009aq, Bergshoeff:2009hq, Gabadadze:2012xv, Sinha:2010ai, Sinha:2010pm, Paulos:2010ke} to the case of arbitrary order $n$ in the curvature tensor. More precisely, we have constructed higher-derivative gravity theories in three dimensions that, on the one hand, $a$) admit simple holographic $c$-theorems and, on the other hand, $b$) exhibit a unique maximally symmetric vacuum. For such theories, we have proven a series of interesting properties, the most salient ones being $i$) the decoupling of ghost gravitational modes about AdS space, $ii$) the enhanced of gauge symmetry at linearized level, $iii$) the presence of slow-decaying mode in AdS, $iv$) the existence of hairy black holes of non-constant curvature that generalize BTZ solution $v$) whose thermodynamics can be rederived from the holographic perspective. The latter is remarkable as it corresponds to a holographic description of black hole microstates for non-supersymmetric solutions of non-constant curvature in higher-derivative theories of arbitrary order in the curvature.

\section*{Acknowledgements}

We thank Jorge Zanelli for discussions. The work of M.C. is partially supported by Mexico's National Council of Science and Technology (CONACyT) grant A1-S-22886, and DGAPA-UNAM grant IN116823. The work of J.M. is supported by FONDECYT Postdoctorado Grant 3230626. The research of R.R. is funded by FONDECYT Postdoctorado Grant 3220663. This work was partially funded by FONDECYT Regular Grants 1221504, 1200022, 1200293, 1210500, 1210635.

\bibliographystyle{JHEP}
\bibliography{Gravities}

\end{document}